\documentclass[preprint,showpacs,epsfig,eqsecnum,aps]{revtex4}
\usepackage{graphicx}
\usepackage{epsfig}
\begin{document}
\title{New results for the missing quantum numbers labeling the quadrupole and
octupole boson basis}

\author{A. Gheorghe$^{(a)}$ and A. A. Raduta$^{a),b)}$}

\address{$^{a)}$Institute of Physics and Nuclear Engineering, Bucharest, POBox MG6, Romania}
\address{
$^{b)}$Department of Theoretical Physics and Mathematics,Bucharest University, POBox MG11,
Romania}

\begin{abstract}
The many $2^k$-pole boson states, 
$|N_kv_k\alpha_k I_kM_k\rangle$ with $k=2,3$, realize the irreducible 
representation (IR)
for the group reduction chains  $SU(2k+1)\supset R_{2k+1}\supset R_3\supset R_2$.
They have been analytically studied and widely used for the description of nuclear systems. However, no analytical expression for the degeneracy
$d^{(k)}_v(I)$ of the $R_{2k+1}$'s IR, determined by the reduction
$R_{2k+1}\supset R_3$, with $k=2,3$ is
available. Thus, the number of distinct values taken by $\alpha_k$ has been 
so far obtained by solving some complex equations. Here we derive
analytical expressions for the degeneracy $d^{(k)}_v(I)$ (k=2,3).
 characterizing the  octupole and quadrupole boson states, respectively.
 The merit of this work consists of the fact that it completes the analytical
 expressions for the $2^k$-pole boson basis for $k=2,3$. The general case of
 $R_{2l+1}$'s IR representation degeneracy is also presented and a compact analytical
 expression for $d^{(l)}_v(I)$ is derived.
\end{abstract}

\maketitle

\section{Introduction}
\label{sec:level1}

Since the liquid drop model was discovered \cite{Bohr}  both
phenomenological and microscopic formalisms use quadrupole and octupole coordinates
 to describe basic properties of
nuclear systems. Based on these coordinates one defines quadrupole and octupole
boson operators in terms of which model Hamiltonians and transition operators  are defined.
Quadrupole properties for a large number  of nuclei can be described by diagonalizing
a quadrupole boson Hamiltonian in the basis $|N_2v_2\alpha_2 I_2M_2\rangle$ associated to the irreducible representation
corresponding to the group reduction chain $SU(5)\supset R_5 \supset R_3\supset R_2$.
The quantum numbers $N_2$ (the number of quadrupole bosons), $v_2$ (seniority)
and $I_2$ are determined
by the eigenvalues of the Casimir operators of the groups $SU(5)$, R$_5$ and R$_3$, respectively.
The angular momentum projection on the axis $z$ is denoted by $M_2$. $\alpha_2$ is usually
 called
the missing quantum number and labels the $R_3$ irreducible representations   which are characterized by the same angular momentum I
and belongs to the same irreducible representation of $R_5$. The name suggests the absence of an intermediate
group between $R_5$ and $R_3$ having a Casimir operator whose eigenvalues would distinguish
the states of the same $I_2$ belonging to the same irreducible representation of $R_5$, v.
The $SU(5)$ boson basis has been analytically derived by three groups, following different
procedures \cite{Cha,Corr,Gh,Ra}. Despite the elegance and the strength  of the methods developed in the above quoted
references, no analytical solution for the number of distinct values
acquired by $\alpha$ for a fixed pair of $v_2$ and $I_2$, denoted by
$d_{v_2}(I_2)$, was presented.
Of course for each $(v_2,I_2)$ one knows to calculate ${\rm d}_{v_2}(I_2)$
numerically, as the number of
solutions ($p_2$) for the inequality:
\begin{equation}
v_2-I_2\le 3p_2\le v_2-\frac{1}{2}(I_2+3r_2),\;\; r_2=\frac{1}{2}\left[1-(-1)^{I_2}\right].
\end{equation}

 Other algorithms for calculating the multiplicity of the irreducible
representations in the chain $SU(5)\supset R_5\supset R_3\supset R_2$ are presented in
Refs. \cite{Kish,Shi,Roho}.

The octupole boson states are classified by the irreducible representations of the groups
involved in the reduction chain $SU(7)\supset R_7 \supset R_3\supset R_2$ and denoted by
$|N_3v_3\alpha_3I_3M_3\rangle$. The quantum numbers are the number of octupole
bosons ($N_3$), seniority ($v_3$),
the missing quantum number ($\alpha_3$), the angular momentum carried by the octupole bosons
($I_3$) and its projection on $z$ axis ($M_3$). The octupole boson number, seniority and angular momentum
are related to the Casimir operator eigenvalues associated to the groups $SU(7)$,
$R_7$ and $R_3$, respectively.
The need of having the octupole boson states degeneracy calculated was first
met in Refs.\cite{Rad1,Rad2}, where a microscopic quadrupole-octupole boson expanded Hamiltonian
was treated in the basis $|N_3v_3\alpha_3I_3M_3\rangle$. Therein the degeneracy
d$_{v_3}(I_3)$ was written as a contour integral which is to be performed each time for a given value
of the pair ($v_3,I_3$), by making use of the Cauchy theorem. Moreover, very useful factorization of
Wigner-Eckart type, for the matrix elements of octupole operators, have
been presented.
Later on a lengthy
recursion equation for d$_{v_3}(I_3)$ was derived in ref.\cite{Shi}. The found equation was
solved numerically for many  ($v_3,I_3$) up to very high values, in Ref.\cite{Roho1}. Extending the
harmonic function method developed in Ref.\cite{Gh}, from the quadrupole bosons to
the octupole bosons, analytical expressions for the states
$|N_3v_3\alpha_3I_3M_3\rangle$  have been derived in Ref.\cite{Roho2}.

The study of octupole degrees of freedom in complex nuclei is an interesting subject which deserves
attention from theoreticians as well as from experimentalists  due to the fact that
systems with static octupole deformations do not exhibit space reflection symmetry
and consequently new specific properties are expected to be found.

It is worth mentioning that the embeding of $R_3$ in $R_7$ mentioned above is
not unique, which results in having several ways of defining a basis for octupole bosons.
The basis mentioned above is used  several groups \cite{Rad1,Rad2,Kamu1,Rogo5}.
The advantage of the provided  basis consists of
that it has formally the same labeling as the quadrupole basis as well as
with the many fermion states in the seniority scheme. It is known the fact
that there is a one to one correspondence between the IR of $R_7$ and those of
$G_2$. This property has been used in Ref.\cite{Roho2} to build a boson basis
$|NvrqsIM\rangle$ with the quantum numbers rqs, named intrinsec quantum numbers,
instead of the missing quantum number $\alpha$.

As mentioned before, in the past the present authors investigated analytically
both the quadrupole and octupole boson basis.  Here we attempt to complete our
previous study
and present analytical expressions for the degeneracy ${\rm d}_{v_3}(I_3)$ and
${\rm d}_{v_2}(I_2)$.
This goal will be touched according to the following plan. In Section 2 we consider
the case of octupole basis while in Section 3 the quadrupole case will be treated.
The general case is treated in Section IV. In the last Section, a short summary will be presented.
\section{Degeneracy of octupole boson states}
\label{sec:level2}
The character of an $R_7$ irreducible representation has the expression
\cite{We1,We2}
\begin{equation}
\chi_{3v}(\varphi_1,\varphi_2,\varphi_3)=\frac{{\rm det}\left(e^{i\varphi_mK_n}-e^{-i\varphi_mK_n}\right)_{m,n=1,2,3}}
{{\rm det}\left(e^{i\varphi_mL_n}-e^{-i\varphi_mL_n}\right)_{m,n=1,2,3}},
\end{equation}
where det$(x_{m,n})_{m,n=1,2,3}$  denotes the determinant associated to the matrix
$(x_{m,n})_{m,n=1,2,3}$. $(L_1,L_2,L_3)$ is the sum of all positive roots for the group
$R_7$, i.e. $(L_1,L_2,L_3)=(5/2,3/2,1/2)$.
The vector $(K_1,K_2,K_3)$ is obtained by adding to $(L_1,L_2,L_3)$  the highest
weight $(S_1,S_2,S_3)$ vector which for the group $R_7$ is equal to $(v,0,0)$.
$(\varphi_1,\varphi_2,\varphi_3)$ is an arbitrary vector.
The restriction of $R_7$ to $R_3$ can be achieved by setting:
\begin{equation}
\frac{\varphi_1}{3}=\frac{\varphi_2}{2}=\varphi_3=\varphi.
\end{equation}
On the other hand the irreducible representation $I$ of the group $R_3$ is characterized by the
character:
\begin{equation}
\chi_I(\varphi)=\frac{\sin{(I+\frac{1}{2})\varphi}}{\sin{\frac{1}{2}\varphi}}.
\end{equation}
Let us consider the set ${\cal C}$ of conjugated elements of $R_3$. The complex functions defined
on ${\cal C}$ can be organized as a Hilbert space ${\cal S}$ with the scalar product defined by:
\begin{equation}
(f,g)=\int^{2\pi}_{0}f^*(\varphi)g(\varphi) \rho(\varphi)d\varphi,
\end{equation}
where $f$ and $g$ are two elements of ${\cal S}$ and $\rho$ denotes the Haar measure for $R_3$ \cite{Vi}
whose expression is:
\begin{equation}
\rho(\varphi)=\frac{1}{\pi}\sin^2\frac{\varphi}{2}.
\end{equation}
The set of functions  $(\chi_I)_I$ is complete in ${\cal S}$ and therefore
any function $\chi_{3v}(\varphi)$ can be expanded as:
\begin{equation}
\chi_{3v}(\varphi)=\sum_{I}{\rm d}^{(3)}_v(I)\chi_I(\varphi)
\end{equation}
The expansion coefficient ${\rm d}^{(3)}_v(I)$ is just the multiplicity of the
representation ($I$)
characterizing the  ($v$) representation splitting. Taking into account that $\chi_I$ are orthonormal
functions one obtains:
\begin{equation}
{\rm d}^{(3)}_v(I)=\int^{2\pi}_{0}\chi^*_I(\varphi)\chi_{3v}(\varphi)\rho(\varphi)d\varphi.
\end{equation}
Changing the integration variable from $\varphi$ to $z=e^{i\varphi}$, the above equation becomes:
\begin{eqnarray}
{\rm d}^{(3)}_v(I)&=&\frac{i}{4\pi}\int_{|z|=1}F(z)dz,\nonumber\\
F(z)&=&\frac{\left( z^{v+1}-1\right) \left(z^{v+2}-1\right) \left(
z^{v+3}-1\right) \left( z^{v+4}-1\right) \left(z^{2v+5}-1\right) \left(
z^{2I+1}-1\right) }{z^{3v+I+2}\left( z^{2}-1\right) \left(z^{3}-1\right)
\left(z^{4}-1\right) \left(z^{5}-1\right) }.
\end{eqnarray}
This expression has been derived by one of us (A.A.R) in Ref.\cite{Rad1}. 
Therein,
results for
several values of $v$ and $I$ ( $0\le I\le 11, 0\le v\le 10$) have been given. 
The nice feature of this expression is that the function $F$ has no pole 
in $z=1$, this value of z being
at a time a zero for numerator. Therefore, it is very easy to be handled
for any pair ($v,I$), by applying the famous residue theorem of Cauchy.
This expression is the starting point for our derivation of an analytical
expression for d$_v(I)$.

We shall touch this goal performing three steps: a) express the fraction
$\left[\left(1-z^2\right)\left(1-z^3\right)\left(1-z^4\right)\left(1-z^5\right)\right]^{-1}$
as a series in $z$ of positive powers; b) separate the singular part, denoted by $G$, from $F$, the holomorphic
rest giving a vanishing contribution to d$^{(3)}_v(I)$; c) calculate the residue for
$G$.

To begin with, let us  calculate the  coefficients for the following expansion,
considered for $|z|<1$:
\begin{equation}
F_{k_{1}k_{2}k_{3}k_{4}}(z)\equiv\frac{1}{\left( 1-z^{k_{1}}\right) \left(
1-z^{k_{2}}\right) \left( 1-z^{k_{3}}\right) \left( 1-z^{k_{4}}\right) }
=\sum\limits_{n=0}^{\infty }N_{k_{1}k_{2}k_{3}k_{4}}(n)\ z^{n}.
\end{equation}
Writing the above series as a product of four series associated to the simple fractions
corresponding to the four factors appearing at denominator in the above equation
one easily obtains that
 $N_{k_{1}k_{2}k_{31}k_{4}}(n)$ is nothing else but the number of solutions $(a,b,c,d)$
of the following equation:
\begin{equation}
k_{1}a+k_{2}b+k_{3}c+k_{4}d=n,
\end{equation}
with $a$, $b$, $c$, $d$ nonnegative integer numbers.
The number of solutions for this equation with four unknown positive integer
numbers can be related to
the number of solutions for an equation having only two nonnegative integer
 unknowns:
\begin{equation}
N_{k_{1}k_{2}k_{3}k_{4}}(n)=\sum
\limits_{r=0}^{n}N_{k_{1}k_{2}}(n-r)N_{k_{3}k_{4}}(r).
\end{equation}
Here $N_{k_{1}k_{2}}(r)$ denotes the number of nonnegative integer numbers solutions ($a,b$)
for the equation
\begin{equation}
k_{1}a+k_{2}b=r.
\end{equation}
For our purposes one needs to know  only the functions with the particular indices
 $(k_1,k_2)=(1,k),\;(k,k+1)$, i.e $N_{1,k}(r)$ and $N_{k,k+1}(r)$.
For the first function $N_{1,k}(r)$, one obviously obtains:
\begin{equation}
N_{1k}(r)=\left[ \frac{r}{k}\right] +1.
\end{equation}

For the other set of  values $k_{1}=k$, $k_{2}=k+1$, Eq.(2.12) becomes:
\begin{equation}
ka+(k+1)b=r,
\end{equation}
which at its turn can be written
\begin{equation}
ku+b=r,
\end{equation}
where $u=a+b$. Taking into account the inequality $0\leq b\leq u$ one obtains:
\begin{equation}
\frac{r}{k+1}\leq x\leq \frac{r}{k},
\end{equation}
and therefore:
\begin{equation}
N_{k,k+1}(r)=\left[ \frac{r}{k}\right] -\left[ \frac{r}{k+1}\right] +\chi
\left( \frac{r}{k+1}\right),
\end{equation}
where $\chi (x)=1$ if $x$ is integer, and $\chi (x)=0$ if $x$ is noninteger.
$\theta(x)$ denotes the step function defined as: $\theta (x)=1$ for 
$x\geq 0$
and  $\theta (x)=0$ for $x<0$. Then, $\chi \left( x\right) =\theta \left( [x]-x\right) $.
In this way analytical expressions for the coefficients $N_{k_1,k_2,k_3,k_4}$ of interest
are obtained:
\begin{eqnarray}
N_{2314}(n)&=&\sum\limits_{r=0}^{n}\left\{ \left[ \frac{r}{3}\right] -\left[
\frac{r}{4}\right] +\chi \left( \frac{r}{4}\right) \right\} \left\{ \left[
\frac{n-r}{2}\right] +1\right\} ,
\nonumber\\
N_{2315}(n)&=&\sum\limits_{r=0}^{n}\left\{ \left[ \frac{r}{2}\right] -\left[
\frac{r}{3}\right] +\chi \left( \frac{r}{3}\right) \right\} \left\{ \left[
\frac{n-r}{5}\right] +1\right\} ,
\nonumber\\
N_{2345}(n)&=&\sum\limits_{r=0}^{n}\left\{ \left[ \frac{r}{2}\right] -\left[
\frac{r}{3}\right] +\chi \left( \frac{r}{3}\right) \right\} \left\{ \left[
\frac{n-r}{4}\right] -\left[ \frac{n-r}{5}\right] +\chi \left( \frac{n-r}{5}
\right) \right\}.
\end{eqnarray}
In what follows we shall use the abbreviations:
\begin{eqnarray}
A(n)&=&\frac{\theta(n)}{2}N_{2314}(n), \nonumber\\
 B(n)&=&\frac{\theta(n)}{2}N_{2315}(n),\nonumber\\
C(n)&=&\frac{\theta(n)}{2}N_{2345}(n).
\end{eqnarray}
Next we take account of the expansion (2.9) for the particular indices
$(k_1,k_2,k_3,k_4)=(2,3,4,5)$ and by brute calculations we write the 
function $F$
as a sum of a holomorphic function, which do not contribute to the integral
(2.8) and a function $G$ having poles in $z=0$. The expression of $G$ is:
\begin{eqnarray}
G(z) &=&\frac{1}{z}\left( \frac{\theta (v-I-3)}{z^{v-I-3}}-\frac{\theta
(v+I-2)}{z^{v+I-2}}+\frac{\theta (I-v-7)}{z^{I-v-7}}\right) F_{2314}(z)\nonumber\\
&&+\frac{1}{z}\left( \frac{\theta (2v+I)}{z^{2v+I}}-\frac{\theta (2v-I-1)}{
z^{2v-I-1}}-\frac{\theta (I-2v-10)}{z^{I-2v-10}}\right) F_{2315}(z)\nonumber\\
&&+\frac{1}{z}\left( \frac{\theta (3v-I)}{z^{3v-I}}-\frac{\theta (3v+I+1)}{
z^{3v+I+1}}+\frac{\theta (I-3v-14)}{z^{I-3v-14}}\right) F_{2345}(z).
\end{eqnarray}
Using this expression, the residue for $F$ is readily obtained and the
final result for
the degeneracy d$^{(3)}_v(I)$, characterizing the $R_7$ irreducible representation, is:
\begin{eqnarray}
{\rm d}^{(3)}_{v}(I) &=&A(v-I-3)-A(v+I-2)+A(I-v-7)+B(2v+I)-B(2v-I-1)\nonumber \\
&&-B(I-2v-10)+C(3v-I)-C(3v+I+1)+C(I-3v-14).
\end{eqnarray}
where the function $A,B$ and $C$ were defined before by Eq.(2.19).
\section{Degeneracy of the quadrupole boson states}
\label{sec:level3}
The case of quadrupole degeneracy may be treated in a similar way with that 
of octupole degeneracy.
Indeed, the character of an irreducible representation is defined by an equation which, formally, is identical to Eq.(2.1), with the difference that now all vectors involved have two components.
Indeed, for $R_5$, $(L_1,L_2)=(3/2,1/2)$, and the highest weight vector is
$(S_1,S_2)=(v,0)$, where $v$ denotes the seniority quantum number for the quadrupole boson system. The reduction from $R_5$ to $R_3$ is achieved by setting
\begin{equation}
\frac{\varphi_1}{2}=\varphi_2\equiv \varphi.
\end{equation}
The final expression for $\chi^{(2)}_v$ can be written as a ratio of two determinants:
\begin{eqnarray}  
\chi^{(2)}_v(\varphi)&=&\frac{\Delta (v)}{\Delta (0)}\;,\rm{where}\nonumber\\
\Delta (v)&=&{\rm det}\left(\matrix{
e^{i\varphi (2v+3)}-e^ {-i\varphi (2v+3)}&  
e^{i\varphi }-e^{-i\varphi }\cr 
e^{i\varphi (v+\frac{3}{2})}-e^{-i\varphi (v+\frac{3}{2})} & e^{
i\frac{1}{2}\varphi }-e^{-i\frac{1}{2}\varphi }}\right).
\end{eqnarray}

The $R_5$ degeneracy caused by the reduction $R_5\supset R_3$ is further expressed as:
\begin{equation}
{\rm d}^{(2)}_v(I)=\int_{0}^{2\pi}\chi^*_I(\varphi)\chi^{(2)}_v(\varphi)\rho(\varphi)d\varphi,
\end{equation}
where $\chi_I$ and $\rho$ are the functions defined by Eqs (2.3) and (2.5), respectively.
Changing the variable $z=e^{i\varphi}$, $d^{(2)}_v(I)$ is expressed as a contour integral:
\begin{equation}
{\rm d}^{(2)}_{v}(I)=\frac{\mathrm{i}}{4\pi }\int\limits_{|z|=1}\frac{1}{z^{2v+I+2}}
\frac{\left( z^{v+1}-1\right) \left( z^{v+2}-1\right) \left(
z^{2v+3}-1\right) (z^{2I+1}-1)}{\left( z^{2}-1\right) \left( z^{3}-1\right) }
\mathrm{d}z.
\end{equation}
Following the same procedure as in the previous section, we perform the expansion

\begin{equation}
F_{k_1k_2}(z)\equiv \frac{1}{(1-z^{k_1})(1-z^{k_2})}=\sum_{n=0}^{\infty}
N_{k_1k_2}(n)z^n.
\end{equation}
The needed expansion coefficients  have been already calculated (see Eqs. (2.13)
, (2.17)):
\begin{eqnarray}
N_{23}(n)&=&\left[ \frac{n}{2}\right] -\left[ \frac{n}{3}\right] +\chi \left(
\frac{n}{3}\right), \nonumber\\ 
N_{13}(n)&=&\left[\frac{n}{3}\right]+1.
\end{eqnarray}

The singular part of the integrand of Eq.(3.4) is:
\begin{eqnarray}
G(z)&=&\frac{1}{z}\left(\frac{\theta(I-2v-5)}{z^{I-2v-5}}+\frac{\theta(2v-I)}{z^{2v-I}}-\frac{\theta(2v+I+1)}{z^{2v+I+1}}\right)F_{13}(z)\nonumber\\
&&+\frac{1}{z}\left(\frac{\theta(v+I)}{z^{v+I}}-\frac{\theta(v-I-1)}{z^{v-I-1}}-\frac{\theta(I-v-3)}{z^{v-I-3}}\right)F_{23}(z).
\end{eqnarray}
With these details the residue for the function $G$ is readily calculated and the final result for multiplicity is:
\begin{eqnarray}
{\rm d}^{(2)}_{v}(I) &=&P(I-2v-5)+P(2v-I)-P(2v+I+1)\nonumber \\
&&+Q(v+I)-Q(v-I-1)-Q(I-v-3).
\end{eqnarray}
where the $Q$ and $P$ denote:
\begin{equation}
Q(n)=\frac{1}{2}\theta (n)N_{13}(n),\quad P(n)=\frac{1}{2}\theta
(n)N_{23}(n).
\end{equation}
Before closing this section we remark that the integral representation for the
$R_7$ and $R_5$ symmetry groups can be written in an unified manner:

\begin{equation}
{\rm d}^{(l)}_{v}(I)=\frac{\mathrm{i}}{4\pi }\int\limits_{|z|=1}\frac{\left(
z^{2I+1}-1\right) \left( z^{2v+2l-1}-1\right)
\prod\limits_{k=1}^{2l-2}\left( z^{v+k}-1\right) }{z^{lv+I+2}\prod
\limits_{k=1}^{2l-2}\left( z^{k+1}-1\right) }\,\mathrm{d}z,
\label{dvl}
\end{equation}
with $l=2$ for $R_5$ and  $l=3$ for $R_7$.

\section{The general case of $R_{2l+1}$ 's degeneracy}
Note that we started by treating first the octupole state degeneracy.
The reason is that the procedure  uses the contour integral  expression for the states
degeneracy derived for the first time \cite{Rad1,Rad2} in connection with the
octupole degrees of freedom. Moreover, since the quadrupole shape coordinates
are widely used for descrybing the collective properties in nuclear systems,
we have applyed the method, formulated in Section II, also to this particular type
of states.
Noteworthy is the fact that the degeneracies for the quadrupole and octupole states
could be written in an unified fashion (\ref{dvl}). This result constitutes a
challenge for us to
prove that Eq. (\ref{dvl}) gives, in fact, the multiplicity for the reduction
$R_{2l+1}\supset R_3$. As a matter of fact this is the objective of the current
section.

For the group $R_{2l+1}$ the character function is :

\begin{equation}
\chi _{lv}(\varphi _{1},\varphi _{2},...,\varphi _{l})=\frac{\mathrm{det}%
\left(e^{i\varphi _{m}K_{n}}-e^{-i\varphi _{m}K_{n}}\right)_{1\leq m,n\leq l}}{\mathrm{%
det}\left(e^{i\varphi _{m}S_{n}}-e^{-i\varphi _{n}S_{m}}\right)_{1\leq m,n\leq l}},
\end{equation}
 where  the vector $K=(K_{1},\ldots ,K_{l})$ is the sum of the root vector
 $L=(L_{1},\ldots ,L_{l})$ and the heighest weight vector
$S=(S_{1},\ldots ,S_{l})$ defined by:
\begin{equation}
L_{k}=l-k+\frac{1}{2},\quad S_{k}=v\delta _{k1},
\end{equation}
The reduction  $R_{2l+1}\supset R_3$ is achieved by the restrictions:
\begin{equation}
\varphi _{k}=k\varphi,;\ k=1,2,..l .
\label{Rest}
\end{equation}

Following the procedure of the previous sections, the degeneracy is defined as
the coefficients of the expansion of $\chi_I(\varphi)$ in terms of $\chi_{lv}(\varphi)$:
\begin{equation}
{\rm d}^{(l)}_{v}=\int_{0}^{2\pi}\chi^*_I(\varphi)\chi_{lv}(\varphi)\rho(\varphi)d\varphi.
\end{equation}
Changing the variable $\varphi$ to  $z=e^{i\varphi }$ one obtains:

\begin{equation}
{\rm d}_{v}^{(l)}(I)=\frac{\mathrm{i}}{4\pi }\int\limits_{|z|=1}\frac{\left(
z^{2I+1}-1\right) U_{l\upsilon }\left( z\right) }{z^{lv+I+2}V_{l\upsilon
}\left( z\right) }\,\mathrm{d}z,
\label{dvlgen}
\end{equation}

where $U_{l\upsilon }$ and $V_{l\upsilon }$ are the following polynomials in z:
\begin{equation}
U_{l\upsilon }\left( z\right) =\left( z^{2v+2l-1}-1\right)
\prod\limits_{k=1}^{2l-2}\left( z^{v+k}-1\right) ,\quad V_{l\upsilon }\left(
z\right) =\prod\limits_{k=1}^{2l-2}\left( z^{k+1}-1\right) ,
\end{equation}
Let us denote by $D_{vl}(m)$ and $N_l(n)$ the coefficients for the $U$ polynomial and
the Taylor expansion associated to $1/V_{lv}$:
\begin{equation}
U_{l\upsilon }\left( z\right) =\sum\limits_{m\geq 0}D_{lv}(m)z^{m},\quad
\frac{1}{V_{l\upsilon }\left( z\right) }=\sum_{n=0}^{\infty }N_{(l)}(n)z^{n}.
\label{expan}
\end{equation}
Here $N_{l}(n)$ denotes the number of solutions $(n_{1},\ldots ,n_{2l-2})$ of
the following equation: 
\begin{equation}
\sum\limits_{k=1}^{2l-2}(k+1)n_{k}=n,
\end{equation}
with $n_{1}$, $\ldots $, $n_{2l-2}$  nonnegative integer numbers.
For the particular cases of $l=2,3$, we have been able to provide analytical
solutions for $N_l(n)$. It is an open question whether this is possible for
the general case.
However, for an arbitrary l, recursive equations for $N_l(n)$ are obtainable.
Note that inserting the expressions of $U$ and $V$ polynomials in
Eq.(\ref{dvlgen}), one obtains the unifying epression (\ref{dvl}) for
the quadrupole and octupole degeneracies.

Inserting the expansion (\ref{expan}) in Eq. (\ref{dvlgen}), the residue can
be easily calculated.
The final expression for $d^{(l)}_v$ is:
\begin{equation}
{\rm d}_{v}^{(l)}(I)=\sum\limits_{n\geq 0}N_l(n)\left[ \theta \left( lv-I-n\right)
D_{lv}(lv-I-n)-\theta \left( lv+I-n+1\right) D_{lv}(lv+I-n+1)\right] .
\label{dvfinal}
\end{equation}

Sumarizing the results obtained in this paper we can asert that the multiplicity
of a irreducible reprezentation $(I)$ of the group $R_3$ in a given ireducible
representation $(v)$ of the group $R_{2l+1}$ is obtained by performing the steps:

a)The character of $(v)$ representation is expanded in terms of the caracters
of IR representations $(I)$.

b)The expansion coefficients are written as contour integral.

c) The contour integral is performed by making use of the Cauchy theorem.

The group reduction $R_{2l+1}\subset R_3$ reflects itself into the restriction
of the character support given by Eq.(\ref{Rest}). Such a reduction is described
in detail in Ref.\cite{We2} for the general case and in Refs. \cite{Kish} and \cite{Rad1}
for the quadrupole and octupole cases, repectively.
A different embedding of $R_3$ into
$R_7$ was proposed in Ref.\cite{Roho1} where the components of the angular
momentum operator, acting on the
space of octupole shape coordinates and the corresponding conjugate momenta,
are expressed as linear combination of the $R_7$ generators. Since the
dimmensions of the irreducible
representations do not depend on the specific realization of the space on which to
group elements act, the degeneracies obtained by the two embeddings are
identical. Actually this can be easilly checked by comparing the results
tabulated in Ref.\cite{Rad1} and \cite{Roho1}, obtained by using different
embeddings.

\section{Summary}
\label{sec:level3}
The main results of the present paper can be summarized as follows:
Based on their integral representations (see Eqs. (2.8) and (3.4)), the $R_7$
and $R_5$ irreducible representation degeneracies are analytically derived and
given by Eqs. (2.21) and (3.8), respectively. The generalization to the group $R_{2l+1}$
has been presented in Section IV and a compact formula for the corresponding degeneracy
$d^{(l)}_v(I)$ was derived (see Eq.(\ref{dvfinal})).

\end{document}